\documentclass[reprint,preprintnumbers,nofootinbib,superscriptaddress,amsmath,amssymb,aps,prd]{revtex4-2}
\pdfoutput=1
\usepackage{graphicx,hyperref,braket,orcidlink, algorithm2e, amssymb, amsmath, dsfont,subfigure,multirow}
\usepackage{lineno}
\graphicspath{{figs/}}
\usepackage{amsmath}
\usepackage{booktabs}
\usepackage{braket}
\usepackage{qcircuit}
\usepackage{bm}

\hypersetup{
	unicode=true,
	pdftoolbar=true,
	pdfmenubar=true,
	pdffitwindow=false,
	pdfnewwindow=true,
	colorlinks=true,
	linkcolor=blue,
	citecolor=magenta,
	filecolor=magenta,
	urlcolor=blue
}

\definecolor{pc1}{rgb}{0.69, 0.25, 0.21}

\usepackage{braket}
\usepackage{dsfont}
\usepackage{nccmath}
\usepackage{scalerel}
\usepackage{bm}
\usepackage{xcolor}
\definecolor{rindou1}{rgb}{0.4431,0.2862,0.7960}
\definecolor{rindou2}{rgb}{0.0078,0.1215,0.4392}
\definecolor{lapis}{rgb}{0.0.0470,0.2941,0.5568}
\definecolor{mn}{rgb}{0.15, 0.35, 0.95}

\begin{document}
\title{Ground state preparation of random all-to-all Hamiltonians using ADAPT-VQE} 


\newcommand{\Hannover}{Institut für Theoretische Physik, Leibniz Universität, Hannover, Germany}
\newcommand{\NCSU}{Department of Physics and Astronomy, North Carolina State University, Raleigh, NC 27607, United States}
\newcommand{\VTPhys}{Department of Physics, Virginia Tech, Blacksburg, VA 24061, United States}
\newcommand{\VTCenter}{Virginia Tech Center for Quantum Information Science and Engineering, Blacksburg, VA 24060, United States}

\author{Sabhyata Gupta\orcidlink{}}
\email{sabhyata.savvy@gmail.com}
\affiliation{\Hannover}

\author{Bharath Sambasivam\orcidlink{}}
\email{bharath.sambasivam@gmail.com}
\affiliation{\VTPhys}
\affiliation{\VTCenter}

\author{Sophia~E. Economou\orcidlink{}}
\affiliation{\VTPhys}
\affiliation{\VTCenter}

\author{Edwin Barnes\orcidlink{}}
\affiliation{\VTPhys}
\affiliation{\VTCenter}

\author{Alexander~F.~Kemper\orcidlink{0000-0002-5426-5181}}
\affiliation{\NCSU}

\author{Raghav G. Jha\orcidlink{0000-0003-2933-0102}}
\email{raghav.govind.jha@gmail.com}
\affiliation{\NCSU}

\begin{abstract}
The ground state of random Hamiltonians with all-to-all interactions such as the quantum Sherrington-Kirkpatrick (SK) model and the Sachdev-Ye-Kitaev (SYK) model follow volume-law entanglement and are expected to be hard to model using tensor networks. In recent years, some progress has been made to push the limit of classical methods using neural quantum states. However, it remains an open question whether there exist quantum algorithms that could offer a quantum advantage over the state-of-the-art classical methods in simulating random Hamiltonians. In this work, we show that one such algorithm, TETRIS-ADAPT-VQE, can construct accurate ground states for dense and sparse SYK models containing up to $N=20$ Majorana fermions achieving fidelities $\geq 99.3\%$ and for the quantum SK model with up to $L=18$ sites achieving fidelities $\geq 99.9998\%$. We find that while the preparation of ground states is efficient (in terms of operator pool size and circuit depth) for the SK model, it is not efficient for either dense or moderately sparse SYK models.

\end{abstract}

\maketitle


\section{Introduction}

It is known that ground states of certain quantum systems can be efficiently explored using classical methods such as tensor networks~\cite{Vidal:2003pmm}. Beyond the tensor networks approach, substantial progress has been made to approximate ground states using neural quantum states (NQSs)~\cite{Carleo:2016svm}. Although, NQSs can approximate ground states with higher entanglement than tensor networks~\cite{Deng:2017uik}, they likely cannot reproduce the correct behavior of \emph{all} volume-law states. For example, it has not yet been possible to efficiently describe the ground state of the SYK model using NQSs~\cite{Passetti:2022ilw, Denis:2023dww, Wurst:2024ajo}. In view of the classical limitations, one is naturally led to considering the preparation of ground states using quantum algorithms---the main motivation of this work. 

\begin{figure}
    \centering
    \includegraphics[width=1.0\linewidth]{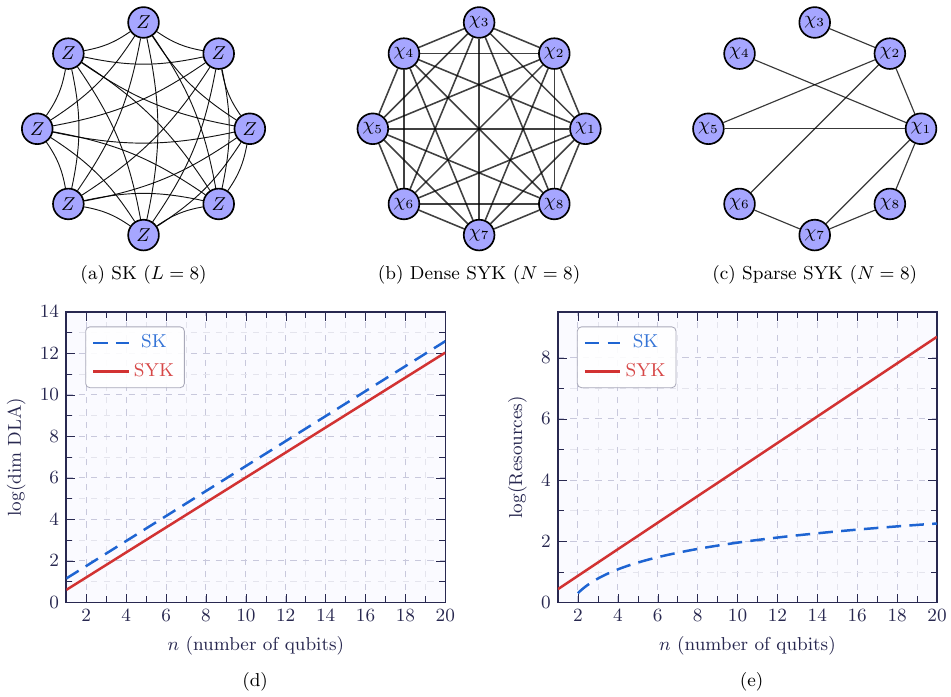}
    \caption{(a)--(c) Schematic representations of the three models we consider in this paper. All of them have all-to-all interactions, and the edges (denoting coupling) are taken at random from some distribution. For the SYK model, two blue circles represent one qubit, while for the SK model, one circle represents one qubit. Panels~(d) and~(e) show the scaling of the dynamical Lie algebra (DLA) dimension and the required resources with number of qubits
$n$ for the SK (dashed blue) and SYK (solid red) models.}
    \label{fig:models_cartoon}
\end{figure}
 
Spin and fermionic models described by random Hamiltonians are relevant for many areas in physics related to spin glasses, non-Fermi liquids, and black hole physics~\cite{Edwards1975,Sherrington1975,  Sachdev_1993, Kitaev:2015, Fujita:2008rs, Luo:2017bno, Gur-Ari:2018okm, Sachdev:2023fim}. It is expected that an improved understanding of these Hamiltonians will eventually turn out to be crucial for explaining several interesting properties of gravitational physics through the gauge/gravity duality on the one hand and for understanding the microscopic features of non-Fermi liquids and disordered systems on the other. The study of random Hamiltonians started with the pioneering work of Edwards and Anderson~\cite{Edwards1975}. They considered a random-disorder version of the nearest-neighbor Ising model to study properties of the spin-glass physics. This was extended by Sherrington and Kirkpatrick~\cite{Sherrington1975} to interactions extending over all spin pairs, as shown schematically in Fig.~\ref{fig:models_cartoon}(a). 
These early investigations paved the way to a broader study of disordered systems, eventually leading to the famous solution by Parisi and to the modern understanding of replica symmetry breaking~\cite{Marinari:1999yb}. Broadly speaking, this framework revealed that spin glasses can possess a highly complex organization of states, with many competing configurations separated by nontrivial energy barriers. The study of these spin-glass models also led to foundational work on Hopfield networks and Boltzmann machines~\cite{Hopfield1982, Amit1985_1, Amit1985_2} and advances in deep learning.

In addition to these applications in condensed matter physics and machine learning problems, in recent years, random-disorder Hamiltonians have attracted renewed interest due to their role in understanding gravitational physics. In particular, the model, known as Sachdev-Ye-Kitaev (SYK)~\cite{Sachdev_1993,Kitaev:2015} has been thoroughly studied. SYK is a quantum-mechanical model in $(0+1)$-dimensions that consists of $N$ randomly interacting Majorana or complex fermions (see Fig.~\ref{fig:models_cartoon}(b) and (c)). In the large $N$ limit, the model has interesting properties in the low-temperature limit $N \gg \beta \mathcal{J} \gg 1$, where $\beta$ is the inverse temperature, while $\mathcal{J}$ is the disorder strength~\cite{Maldacena:2016hyu}. In this low-temperature limit, it is maximally chaotic and saturates the chaos bound for the Lyapunov exponent~\cite{MaldacenaJHEP2016}. Due to its relevance to quantum gravitational physics and quantum chaos, this model has been thoroughly studied using classical~\cite{Kobrin:2020xms, Xu:2022vko} and quantum computing methods~\cite{Babbush:2018mlj,Asaduzzaman:2023wtd,Jha:2024vcw,Granet:2025yys,Chowdhury:2025loy}. Apart from the real-time dynamics of the dense~\cite{Asaduzzaman:2023wtd, Chen:2025xpq} and sparse SYK models~\cite{Chen:2025xpq, Granet:2025yys}, the preparation of thermofield double (TFD) states~\cite{Su:2020zgc} and thermal Gibbs states~\cite{Araz:2024xkw, Kundu:2025xxw} has also been studied using different quantum algorithms for up to $N = 16$ Majorana fermions. However, to our knowledge, no work has focused on the preparation of ground states of the SYK model using variational quantum algorithms. 

One of the difficulties in exploring the random Hamiltonians considered in this work stems from the non-local (all-to-all) structure of the Hamiltonian, making it difficult to construct and optimize parametrized quantum circuits that are expressive enough to reproduce the highly entangled ground states of these models. This can be understood from the corresponding rapid growth of the dynamical Lie algebras of these models (see Fig.~\ref{fig:models_cartoon}(d) and (e)). We tackle these issues using the Adaptive Derivative Assembled Problem-Tailored Variational Quantum Eigensolver (ADAPT-VQE)~\cite{Grimsley:2018wnd}, which has been shown to produce compact circuits~\cite{Ramoa:2024nno,Ramoa:2024zed,AnastasiouPRR2024} while avoiding optimization issues that plague fixed variational algorithms with fixed ans\"{a}tze~\cite{Grimsley:2022azc}. It was argued in Ref.~\cite{Chen2024} using quantum phase estimation that the ground states of sufficiently sparse random Hamiltonians (such as the quantum SK and SYK models) might be quantum mechanically easy to prepare. Specifically, in Ref.~\cite{Herasymenko:2022wup}, it was argued that for $d$-sparse Hamiltonians where $d = O(1)$, the preparation of ground states is easy using quantum phase estimation. However, practical demonstrations on near-term or early-fault tolerant devices are missing. 

In this paper, we take a step towards this goal by investigating whether it is possible to variationally prepare the ground states  of three random Hamiltonians---the quantum SK model and two versions (dense and sparse) of the SYK model---using compact, hardware-friendly circuits. The variational quantum algorithm we use is the TETRIS version of ADAPT-VQE~\cite{AnastasiouPRR2024}, which has been shown to produce compact state-preparation circuits for other types of simulation problems such as molecular systems. We perform noiseless classical simulations to numerically evaluate the scaling of the circuit complexity of preparing the ground states of both the SK and SYK models.
For the quantum SK model, we construct ground states efficiently with polynomially scaling two-qubit gate depth for up to $L=18$ qubits with a fidelity exceeding 99.9998\% in all cases. For the dense and sparse SYK models, we prepare ground states for up to $N=20$ Majorana fermions with fidelity always greater than $99.3\%$. However, we find that the two-qubit gate depth scales exponentially with the system size for both the dense and sparse SYK models. The choice of a symmetry-informed Pauli operator pool for both the quantum SK and SYK model is crucial to achieve this performance. 

The outline of this paper is as follows. In Sec.~\ref{sec:SK}, we consider the quantum SK model and obtain ground states for a system size of up to 18 qubits with very high fidelity. Then, in Sec.~\ref{sec:SYK}, we present the results for the ground-state preparation and resources required for the dense and sparse SYK models containing up to $N=20$ Majorana fermions (10 qubits).  We end the paper with a summary and possible extensions of our work. 


\section{\label{sec:SK}Quantum Sherrington-Kirkpatrick model}

\begin{figure}
    \centering
\hspace*{-0.8cm}\includegraphics[width=0.95\linewidth]{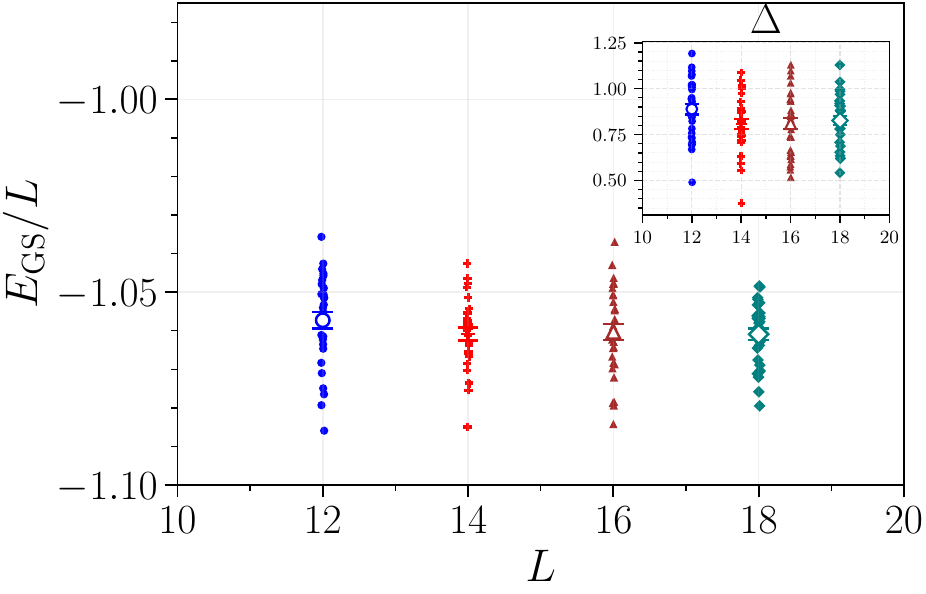}
    \caption{The exact diagonalization results for the ground state energy and the gap $\Delta = E_{1} - E_0$ (inset) for the quantum SK model with up to 18 qubits and $\Gamma = 1$. The results are averaged over 30 instances of the model. The gap for the SK model is significantly larger than that for the SYK model shown in the inset of Fig.~\ref{fig:syk_exact_gs_gap}.}
    \label{fig:sk_exact_gs_gap}
\end{figure}

In this section, we consider the quantum SK model~\cite{Sherrington1975}, with  Hamiltonian 
\begin{equation}
    H = -\sum_{1\le i<j\le L} J_{ij} Z_{i}Z_
    {j} -\Gamma \sum_{i=1}^{L} X_{i}. 
\end{equation}
The random couplings $J_{ij}$ are taken from a Gaussian distribution with zero mean and variance $\mathcal{J}/N$. We set $\mathcal{J} = 1$.  
This model has been extensively studied and is relevant for spin-glass physics, which shows volume-law entanglement in the ground state. Specifically, there is a phase transition at $\Gamma/\mathcal{J} = 1.5$~\cite{Schindler2022}, with a spin-glass phase for $\Gamma \to 0$ and a paramagnetic phase for $\Gamma \gg 1$.
This model is an extension of the earlier Edwards-Anderson (EA) model~\cite{Edwards1975} that includes all-to-all interactions instead of just nearest-neighbor disordered interactions. The first computations of thermodynamic properties using the replica symmetric ansatz lead to unphysical results such as negative entropy at zero temperature~\cite{Edwards1975}. Later, the exact ground-state energy for a particular limit of the model ($\Gamma = 0$) was calculated by Parisi~\cite{Parisi1979, Parisi1980, Aizenman:1987gp} using ideas related to spontaneous replica symmetry breaking~\cite{Mezard1987}. We show results from exact diagonalization for the ground-state energy and the spectral gap $(\Delta = E_{1}-E_{0}$) for the quantum SK model averaged over 30 instances in Fig.~\ref{fig:sk_exact_gs_gap}. Although the Parisi solution gives the exact energy (with no transverse field), it does not provide a method for finding the ground state. 
Over the past five decades, many attempts have been made to approximate the ground state in various limits, including classical methods such as neural quantum states (NQSs)~\cite{Schindler2022, Wurst:2024ajo} and QAOA methods for the classical SK model (in the absence of a transverse field) ~\cite{Farhi:2019xsx, Boulebnane:2025grd}, with reasonable success. Solving the quantum SK model with quantum methods requires a more general variational approach beyond QAOA owing to the volume-law entanglement in the ground state.

\begin{table}[h!]
\centering
\setlength{\tabcolsep}{6pt} 
\renewcommand{\arraystretch}{1.2} 
\begin{tabular}{|c|c|c|c|c|}
\hline
$L$ & Iterations & Parameters & $\delta$ (in $\%$) & Fidelity \\ 
\hline 
12 & $25(2)$ & $118(13)$ & $0.0036(3) \%$ & 0.999943(5) \\ \hline  
14 & $30(1)$ & $174(8)$ & $0.0048(4) \%$ & 0.999908(8) \\ \hline  
16 & $35(2)$ & $220(15)$ & $0.0059(4) \%$ & 0.999868(10) \\  \hline  
18 & $35(3)$ & $253(17)$ & $0.0068(4) \%$ & 0.999828(13) \\  \hline  
\hline
\end{tabular}
\caption{The number of ADAPT iterations, variational parameters, ground-state energy error $\delta$, and state fidelity relative to exact diagonalization for the SK model with $L$ sites and $\Gamma = 1$ averaged over five instances of the model couplings.}
\label{tab:scaling_SK_Gamma1}
\end{table}

\begin{table}[h!]
\centering
\setlength{\tabcolsep}{5pt} 
\renewcommand{\arraystretch}{1.2} 
\begin{tabular}{|c|c|c|c|c|}
\hline
$L$ & Iterations & Parameters & $\delta$ (in $\%$) & Fidelity \\ 
\hline 
12 & $14(1)$ & $76(1)$ & $0.000036(8) \%$ & 0.999999(0) \\ \hline  
14 & $16(1)$ & $103(1)$ & $0.000032(3) \%$ & 0.999999(0) \\ \hline  
16 & $18(1)$ & $135(1)$ & $0.000057(11) \%$ & 0.999999(0) \\  \hline  
18 & $20(1)$ & $170(1)$ & $0.000033(2) \%$ & 0.999999(0) \\  \hline  
\hline
\end{tabular}
\caption{The number of ADAPT iterations, variational parameters, ground-state energy error $\delta$, and state fidelity relative to exact diagonalization for the SK model with $L$ sites and $\Gamma = 3$ averaged over five instances of the model couplings.}
\label{tab:scaling_SK_Gamma3}
\end{table}

\begin{figure}
    \centering
\includegraphics[width=0.95\linewidth]{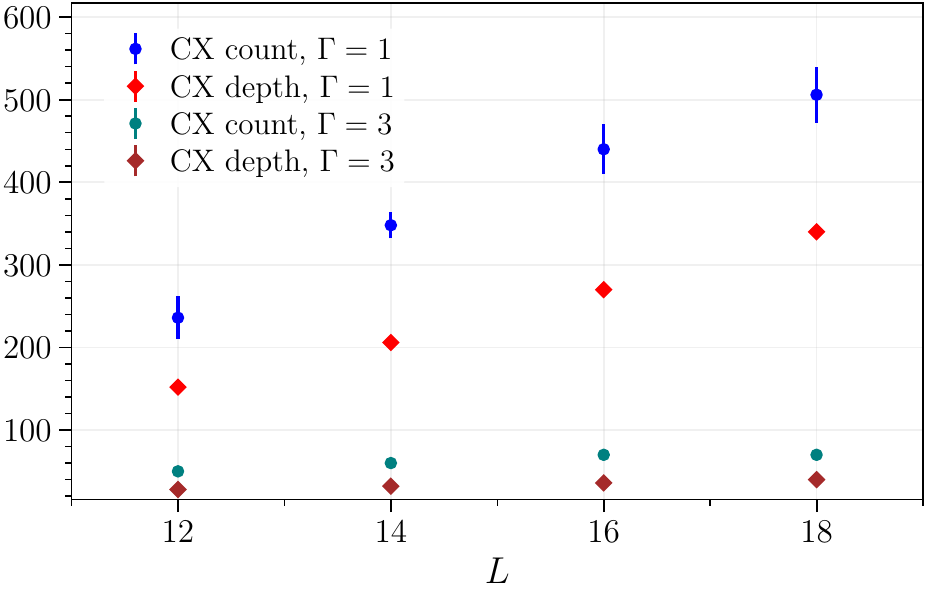}
    \caption{The CX count and depth of the unitary circuit that prepares the ground state averaged over 5 instances of the quantum SK model with $L$ sites. Results are shown for two different values of the transverse magnetic field, $\Gamma$.}
    \label{fig:sk_CNOT}
\end{figure}

In order to simulate this problem using ADAPT-VQE, we first construct a pool of operators that satisfy the symmetries of the Hamiltonian. The algorithm will use this pool to iteratively construct the state-preparation circuit. A brief review of ADAPT-VQE is given in Appendix~\ref{sec:AVQE}. Starting with a parent pool consisting of all one- and two-weight Pauli strings, we remove operators that do not commute with symmetries of the system. This is because ADAPT-VQE never selects these symmetry-breaking operators if the reference state belongs to the same symmetry sector as the ground state, because in this case the cost-function gradients that the algorithm uses to determine which operators to insert into the circuit vanish identically. The SK model possesses time-reversal and parity symmetries. The time-reversal symmetry operator $\mathcal{T}$, being anti-unitary, acts on $-iH$ as
\begin{equation}
  \mathcal{T}(-i H)\mathcal{T}^{-1}
  = (-i)^{*}\,\mathcal{T} H \mathcal{T}^{-1}
  = i\,\mathcal{T} H \mathcal{T}^{-1}.
\end{equation}
We take $\mathcal{T}$ to satisfy $\mathcal{T}^2 = +1$. Its action on single-site
Pauli operators is
\begin{align}
  \mathcal{T} X_i \mathcal{T}^{-1} &= X_i, \qquad
  \mathcal{T} Z_i \mathcal{T}^{-1} = Z_i, \qquad
  \mathcal{T} Y_i \mathcal{T}^{-1} = -\,Y_i,
\end{align}
which corresponds to complex conjugation in the computational basis (since $X$ and $Z$ are real, while $Y$ is purely imaginary). The parity symmetry is described by the operator $\mathcal{P}_X=\prod_{i=1}^LX_i.$

The quantum SK Hamiltonian, built from random $Z_i Z_j$ couplings and transverse $X_i$ fields with real coefficients, is manifestly invariant under $\mathcal{T}$ and $\mathcal{P}_X$. The ground state of models with such symmetries can be represented using a purely real wavefunction. Therefore, a nontrivial variational ansatz 
of the form $\exp(-i\theta P)$ requires pool operators $P$ that \emph{only} have an odd number of $Y$ operators. This restriction, combined with the required commutation with $\mathcal{P}_X$, leaves two classes of valid operators in the pool:
\begin{equation}
    \mathbf{P}_{\text{SK}}\equiv\left\{Z_iY_j,\,Y_iZ_j\,\vert\,i<j\right\},
\end{equation}
leading to a total of $O(L^2)$ operators in the pool. Note that there are no single-qubit generators in this operator pool.
We use an equal superposition of N\'eel states as the reference state for ADAPT-VQE in all our simulations; this state is an eigenstate of both symmetry operators. 

The key results from our simulations of the TETRIS-ADAPT-VQE algorithm applied to the SK model are summarized in Tables~\ref{tab:scaling_SK_Gamma1} and 
\ref{tab:scaling_SK_Gamma3}.
We quantify the energy accuracy using the relative ground-state energy error $\delta =
    \left|
    \frac{E_{\mathrm{GS}}-E_{\mathrm{ADAPT}}}{E_{\mathrm{GS}}}
    \right|$, while we gauge the accuracy of the prepared ground state using the fidelity relative to the state obtained from exact diagonalization. Tables~\ref{tab:scaling_SK_Gamma1} and 
\ref{tab:scaling_SK_Gamma3} show results for $\Gamma=1$ and $\Gamma=3$, respectively.
Due to the change in the structure of the ground state across the phase transition at $\Gamma=1.5$, we expect the complexity to prepare it to vary between $\Gamma = 1$ and $\Gamma = 3$, and this is evident in Tables~\ref{tab:scaling_SK_Gamma1} and 
\ref{tab:scaling_SK_Gamma3}, where $\Gamma =3$ requires a consistently lower number of iterations and operators to achieve similar fidelities, since it is further away from the phase boundary at $\Gamma = 1.5$ compared to $\Gamma = 1$. We show the \texttt{CNOT} depth and count for $\Gamma = 1,3$ for the quantum SK model averaged over five instances of the couplings in Fig.~\ref{fig:sk_CNOT}. 

\section{\label{sec:SYK}Dense and sparse SYK models}

SYK models are disordered fermionic many-body systems. They are an extension of an earlier model by Sachdev and Ye for quantum Heisenberg magnets with Gaussian infinite-range couplings~\cite{Sachdev:1992fk}. 
In 2015, Kitaev showed~\cite{Kitaev:2015} that a similar fermionic model with quartic Majorana interactions at large $N$ yields a model with emergent low-temperature properties that indicate a connection to holography. The SYK model is a 0+1-dimensional model of fermions (either Majorana or complex fermions). In this work, we will consider Majorana fermions $\chi$. They obey the standard anti-commutation relation: $\{\chi_{a}, \chi_{b}\} = 2\delta_{ab}$. The dense SYK (dSYK) Hamiltonian is given by
\begin{align}
    \mathcal{H}_{\text{dSYK}} = \sum_{1\le i<j<k<l\le N} J_{ijkl} \,\chi_{i} \chi_{j} \chi_{k} \chi_{l}, 
\label{eq:SYK_dense}
\end{align}
where $N$ denotes the number of Majorana fermions, and the random couplings $J_{ijkl}$ are taken from a Gaussian distribution with mean $\overline{J_{ijkl}}=0$ and variance equal to $\overline{J_{ijkl}^{2}} = \frac{3! \mathcal{J}^2}{N^3}$. We work in units where $\mathcal{J}=1$ for convenience. Each term in the Hamiltonian given in Eq.~\eqref{eq:SYK_dense} is a four-fermion interaction (usually called $q=4$ SYK in the literature). Generalizations for higher-body interactions ($q>4$) exist ~\cite{Maldacena:2016hyu}, but will not be considered in this work. In order to map the Hamiltonian onto qubits, we note that two Majorana (real) fermions can be described by one complex fermion, and therefore we need $n = N/2$ qubits to encode the Hamiltonian. We will reserve $N$ for the number of Majorana fermions and $n$ for the number of qubits in this paper. We follow the standard approach of using the Jordan-Wigner transformation to rewrite $\chi$'s in terms of Pauli matrices, $X, Y, Z$ as:
\begin{align}
\chi_{2k-1} &= \Big(\prod_{j=1}^{k-1} Z_{j}\Big) \otimes X_{k}  \otimes \mathbb{I}^{\otimes (N -2k)/2},  \nonumber \\
\chi_{2k} &= \Big(\prod_{j=1}^{k-1} Z_{j}\Big) \otimes Y_{k} \otimes \mathbb{I}^{\otimes (N -2k)/2}.
\label{eq:JWtransform}
\end{align}

In addition to the standard SYK model described by Eq.~\eqref{eq:SYK_dense}, we also consider the sparsified version of the model (sSYK) where the number of terms in the Hamiltonian is reduced from $\mathcal{O}(N^4)$ to $\mathcal{O}(N)$~\cite{Garcia-Garcia:2020cdo, Xu:2020shn, Jha:2024vcw}. The Hamiltonian for the sparse SYK model is given by:
\begin{align}
    \mathcal{H}_{\text{sSYK}} = \sum_{1\le i<j<k<l\le N} J_{ijkl} \,\mathcal{P}_{ijkl} \,\chi_{i} \chi_{j} \chi_{k} \chi_{l}, 
\label{eq:SYK_sparse}
\end{align}
where $\mathcal{P}_{ijkl}$ is set to $0$ or $1$ depending on whether a random number chosen randomly from $[0,1)$ is more or less than $p = 24k_{\text{s}}/(N^3)$, where $k_{\text{s}}$ is the sparsification parameter. The dense model is obtained when we take $k_{\text{s}} = N^3/24$, corresponding to $p = 1$, i.e., keeping all the terms in the Hamiltonian. For a given value of 
$k_{\text{s}}$, the SYK Hamiltonian will have $k_{\text{s}}N$ terms on average. 
The critical sparsity required for holographic interpretations of the SYK-type model has been the subject of many investigations. The Lyapunov exponent was found to be relatively insensitive to sparsity above the percolation threshold~\cite{Garcia-Garcia:2023jlu}, which corresponds to $k_s \approx 1$. In this work, we focus on the sparse SYK model with fixed $k_s=9$, since this limit has been shown to be `safely' in the holographic regime~\cite{Orman:2024mpw, Jha:2024nbl}.


\subsection{Ground-state preparation and entanglement entropy}

We now consider the problem of how effectively ADAPT-VQE can prepare the ground state of the SYK-type Hamiltonian. We will consider both the dense and sparse models for up to ten qubits, corresponding to $N=20$ Majorana fermions. For the sparse variant, we consider $k_s=9$.

For ADAPT-VQE, we construct a pool of operators that satisfy the symmetries of the Hamiltonian. Starting again with a parent pool consisting of all one-and two-weight Pauli strings, we remove operators that do not commute with symmetries of the system. The SYK Hamiltonian (both dense and sparse) has a parity symmetry, i.e., $[H, \mathcal{P}_Z] = 0$ where 
$\mathcal{P}  = \prod_{i=1}^{N/2} Z_i$.
Therefore, it is useful to construct the pool out of operators that commute with $\mathcal{P}$. There are four classes of operators that are left in the parent pool after the pruning based on the above parity symmetry:
\begin{equation}
    \mathbf{P}_{\text{SYK}}\equiv \left\{Z_i,\,X_iY_j,\,Y_iX_j,\,Z_iZ_j\,\vert \,i<j\right\}
\end{equation}
The size of this operator pool scales as $O(n^2)$, where $n=N/2$ is the number of qubits in the model. We use the N\'{e}el state $|01\rangle^{\otimes n/2}$ or $\ket{10}^{\otimes n/2}$ belonging to the correct $\mathbb{Z}_2$ parity sector of the SYK Hamiltonian as the reference state in all our ADAPT-VQE simulations.

\begin{figure}
    \centering
    \hspace*{-0.8cm}
    \includegraphics[width=0.95\linewidth]{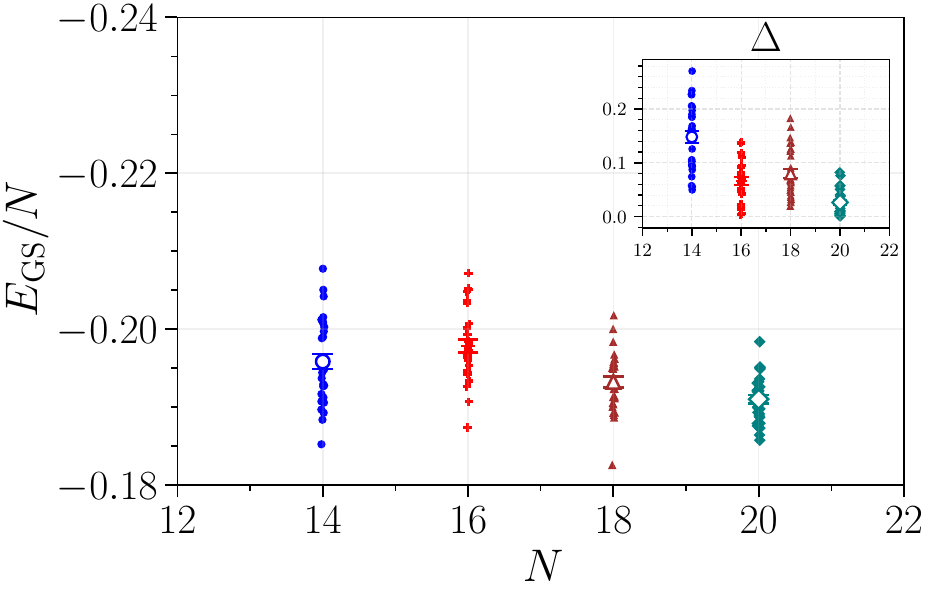}
    \caption{
    Exact diagonalization results for the ground-state energy and spectral
    gap $\Delta$ of the dense SYK model up to $N=20$. The results are
    averaged over $30$ disorder realizations.
    }
    \label{fig:syk_exact_gs_gap}
\end{figure}

As a first diagnostic of the ansatz complexity, we compute exact diagonalization results for the ground-state energy and spectral gap of the dense SYK model. The results, averaged over $30$ disorder realizations, are shown in Fig.~\ref{fig:syk_exact_gs_gap}. The SYK spectral gap $\Delta$ is considerably smaller than that of the quantum SK model shown in the inset of Fig.~\ref{fig:sk_exact_gs_gap}, indicating that the low-energy spectrum is more compressed and therefore more difficult to resolve variationally.

\begin{figure}
    \centering
    \includegraphics[width=0.95\linewidth]{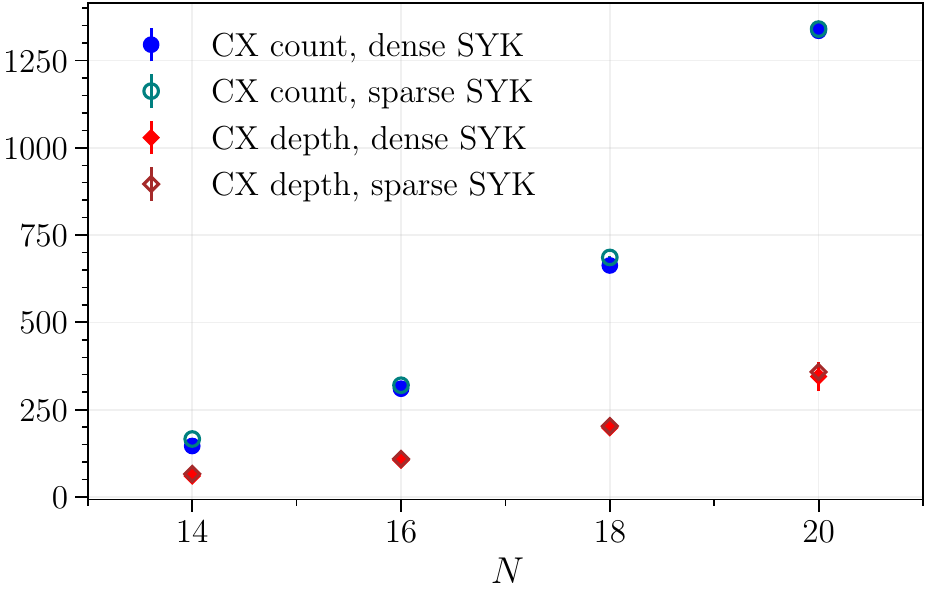}
    \caption{
    \texttt{CNOT} count and depth of the ADAPT-VQE circuits that prepare the ground state, averaged over five disorder realizations of the dense and sparse SYK models.
    }
    \label{fig:syk_CNOT}
\end{figure}

We again quantify the ADAPT-VQE energy accuracy using the relative ground-state energy error $\delta =
    \left|
    \frac{E_{\mathrm{GS}}-E_{\mathrm{ADAPT}}}{E_{\mathrm{GS}}}
    \right|$, while the accuracy of the prepared state is quantified by the fidelity relative to the ground state obtained from exact diagonalization. 
The results for the dense and sparse SYK models are shown in
Tables~\ref{tab:scaling-denseSYK} and~\ref{tab:scaling-sparseSYK}, respectively. In both cases, ADAPT-VQE reaches high fidelity for $N\leq 20$, but the number of iterations and parameters grows rapidly with system size. The scaling of the resulting \texttt{CNOT} count and depth is shown in Fig.~\ref{fig:syk_CNOT}, and appears to be consistent with an exponential growth or a higher-degree polynomial over the system sizes studied.

Although sparsification substantially reduces the number of Pauli terms in the Hamiltonian, we find that it does not lead to a comparable reduction in the variational complexity required for ground-state preparation. For example, with $N=20$, the dense SYK Hamiltonian contains $4845$ Pauli terms, whereas the sparse $k_s=9$ Hamiltonian contains approximately $180$ Pauli terms on average. Thus, sparsification removes roughly $96\%$ of the Pauli terms. This reduction is important for the practical measurement cost because fewer Hamiltonian terms need to be estimated. However, in the exact state-vector simulations considered here, where measurement noise is absent, the dense and sparse models display very similar ADAPT-VQE convergence behavior. In particular, the number of ADAPT iterations, number of variational parameters, energy accuracy, fidelity, and circuit depth are all comparable for the two models.

This similarity can be understood from the structure of the ground state rather than from the number of Hamiltonian terms alone. For both dense SYK and sparse SYK with $k_s=9$, the ground state remains highly entangled and exhibits volume-law entanglement. Therefore, although the sparse Hamiltonian is much cheaper to measure, its ground state is not significantly easier to prepare variationally. Only when the model is sparsified much more strongly, roughly in the regime $k_s \lesssim 1$, does Hilbert-space fragmentation set
in, and the Hamiltonian decomposes into disconnected blocks. In that limit, one expects the preparation problem to change qualitatively. This fragmented regime has also been discussed in the context of quantum chaos in Ref.~\cite{Iizuka:2024die}.

\begin{table}[h!]
\centering
\setlength{\tabcolsep}{6pt} 
\renewcommand{\arraystretch}{1.2} %
\begin{tabular}{|c|c|c|c|c|}
\hline
$N$ & Iterations & Parameters & $\delta$ (in $\%$) & Fidelity \\ 
\hline 
14 & $31(1)$ & $124(1)$ & $0.16(11) \%$ & 1.0(0) \\ \hline  
16 & $54(1)$ & $250(2)$ & $0.11(2) \%$ & 0.9993(1) \\  \hline  
18 & $100(2)$ & $504(2)$ & $0.09(3) \%$ & 0.9982(6) \\  \hline  
20 & $172(1)$ & $974(3)$ & $0.31(9) \%$ & 0.9936(27) \\  \hline  
\hline
\end{tabular}
\caption{\label{tab:scaling-denseSYK}The number of ADAPT iterations, variational parameters, ground-state energy error $\delta$, and state fidelity relative to exact diagonalization for the dense SYK model with $N$ Majorana fermions averaged over five instances of the model.}
\end{table}

\begin{table}[h!]
\centering
\setlength{\tabcolsep}{6pt} 
\renewcommand{\arraystretch}{1.2} %
\begin{tabular}{|c|c|c|c|c|}
\hline
$N$ & Iterations & Parameters & $\delta$ (in $\%$) & Fidelity \\ 
\hline 
14 & $33(1)$ & $126(1)$ & $0.06(3) \%$ & 0.99965(3) \\ \hline  
16 & $45(4)$ & $209(18)$ & $0.78(43) \%$ & 0.995(4) \\  \hline  
18 & $99(2)$ & $499(1)$ & $0.50(16) \%$ & 0.997(1) \\  \hline  
20 & $177(4)$ & $980(13)$ & $0.55(13) \%$ & 0.9966(8) \\  \hline \hline 
\end{tabular}
\caption{\label{tab:scaling-sparseSYK}The number of ADAPT iterations, variational parameters, ground-state energy error $\delta$, and state fidelity relative to exact diagonalization with $N$ Majorana fermions for the sparse $k_s=9$ SYK model averaged over five instances of the model.}
\end{table}

\begin{figure}[h!]
    \centering
    \hspace*{-0.8cm}
    \includegraphics[width=0.95\linewidth]{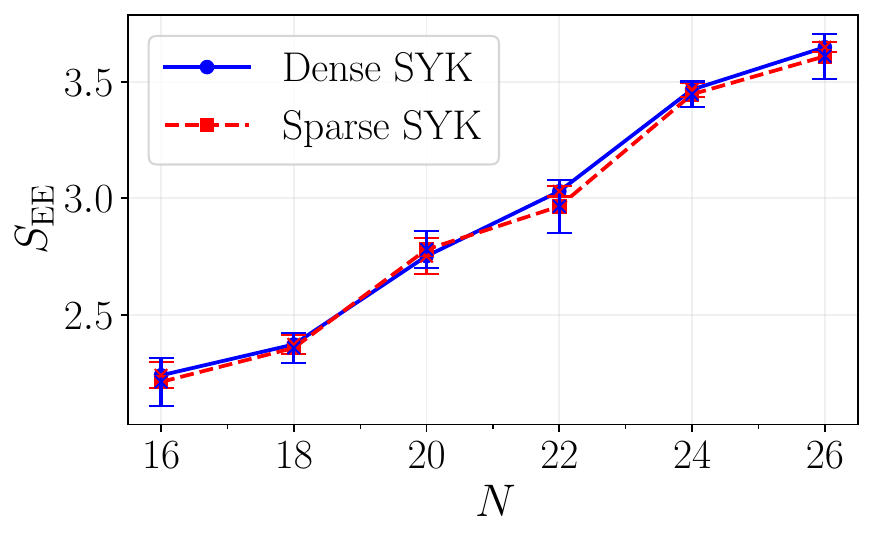}
    \caption{
    Bipartite von Neumann entanglement entropy for the dense SYK model and the sparse SYK model with $k_s=9$, for $N\in[16,26]$. Both models show volume-law scaling, $S_{\mathrm{EE}}\propto N$, with similar slopes. The
    results are averaged over ten disorder realizations of both models. 
    }
    \label{fig:vn_entropy}
\end{figure}

To further understand why the sparse model remains difficult to prepare, we compare the bipartite von Neumann entanglement entropy of the dense and sparse SYK ground states. For a given $N$, we first compute the normalized ground state $\ket{\psi_0}$ of the SYK Hamiltonian. The Hilbert-space dimension is $2^{N/2}$. We then choose a bipartition $\mathcal H=\mathcal H_A\otimes\mathcal H_B$ and trace out subsystem $B$, obtaining
\begin{equation}
    \rho_A
    =
    \mathrm{Tr}_B
    \left(
    \ket{\psi_0}\bra{\psi_0}
    \right).
\end{equation}
The bipartite von Neumann entanglement entropy is then
\begin{equation}
    S_{\mathrm{EE}}
    =
    -\mathrm{Tr}
    \left(
    \rho_A \ln \rho_A
    \right).
\end{equation}
The results for the entanglement entropies are shown in Fig.~\ref{fig:vn_entropy} and Table~\ref{tab:syk_dense_sparse}. Both dense SYK and sparse SYK with $k_s=9$ display volume-law scaling, $S_{\mathrm{EE}}\propto N$, with nearly identical coefficients. This provides direct evidence that sparsification at $k_s=9$ reduces the Hamiltonian measurement cost but does not substantially reduce the complexity of the ground-state wave function.

\renewcommand{\arraystretch}{1.4} 
\begin{table}[h!]
\centering
\begin{tabular}{c|c|c}
\toprule
$N$ & Dense SYK & Sparse-SYK, $k_s=9$ \\ \hline
\hline
16 & 2.24(6) & 2.21(10) \\
\hline 
18 & 2.37(4) & 2.36(7) \\
\hline 
20 & 2.75(8) & 2.78(8) \\
\hline 
22 & 3.03(2) & 2.97(11) \\
\hline 
24 & 3.47(3) & 3.47(6) \\
\hline 
26 & 3.65(2) & 3.61(10) \\
\hline
\end{tabular}
\caption{von Neumann entanglement entropies of dense and sparse ($k_s=9$) SYK models for different numbers $N$ of Majorana fermions. The comparable entropies of the two models signal that their ground states are comparably entangled.}
\label{tab:syk_dense_sparse}
\end{table}

This reported behavior is consistent with previous studies of the four-fermion SYK model. Exact diagonalization calculations up to $N=44$ Majorana fermions find that, for an equal bipartition with subsystem size $N/2$, the ground-state entanglement entropy scales approximately as
$S_{\mathrm{EE}} \simeq 0.169\, N$ up to an additive constant. This confirms the volume-law character of the ground state. The coefficient changes only weakly with increasing weight of the fermionic interactions ($q$) in the Hamiltonian and approaches $\frac{\ln 2}{4}\,N \simeq 0.173\,N$ in the large-$q$ limit~\cite{Liu:2017kfa,Huang:2017nox,Zhang:2020kia,Zhang:2022yaw}. Thus, from the perspective of entanglement, the difficulty of preparing the four-fermion  SYK ground state is not expected to be qualitatively different from that of higher-$q$ SYK models. This supports the interpretation that the comparable
ADAPT-VQE performance for dense and sparse $k_s=9$ SYK is controlled primarily by the entanglement structure of the ground state, rather than by the raw number of Pauli terms in the Hamiltonian, and if an efficient algorithm can be found for $q=4$, it will likely also work for larger $q$.

For completeness, we also checked whether the scalings of circuit complexity we are seeing are a consequence of the choice of reference state and operator pool. First, rather than restricting ourselves to the unentangled N'{e}el states as the reference state, we considered highly entangled Clifford states as initial states. We also generalized our operator pool to include higher-weight Pauli operators to see if the convergence and scaling of circuit size for the preparation of SYK ground states can be improved. Additionally, we tested a Majorana operator-based pool built directly from the SYK interaction structure. Specifically, we generated all four-body Majorana operators $\chi_i \chi_j \chi_k \chi_l$ with $i < j < k < l$, applied the Jordan-Wigner encoding, and retained only those operators satisfying the parity symmetry constraint $[\mathcal{P}, \mathcal{O}] = 0$ where $\mathcal{P} = \prod_{i=1}^{n} Z_i$. We considered operators with maximum Pauli weight 3 appended to the usual weight-2 Pauli strings used in the main results, and used the TETRIS variant of ADAPT-VQE~\cite{AnastasiouPRR2024} for all runs. The weight-3 pool achieved comparable convergence and fidelity to the main results, but at approximately $3\times$ higher CX gate cost. We therefore conclude that the weight-2 pool used in the main text is the more resource-efficient choice. We found that neither alternative initial states nor the addition of higher-weight Pauli operators to the pool improved the resource scaling, confirming that the exponential scaling is likely intrinsic to the SYK model and not an artifact of hyperparameter choices in the ADAPT-VQE algorithm. We provide additional details on the reference state comparison in Appendix~\ref{app:robustness}.


\section{Summary and Conclusion}

In this work, we considered the preparation of volume-law-entangled ground states of three models: the SK model and dense and sparse versions of the SYK model. For the SK model, we consider up to $L=18$ (18 qubits), and for the SYK models, we explored up to $N=20$ Majorana fermions (10 qubits). Our results clearly show that the growth of circuit depth and parameter scaling is polynomial for the SK model, while it appears to be exponential for the dense and sparse SYK model. Hence, while the state preparation for the SK model may be a suitable task for a quantum computer, it appears to be difficult for the SYK-type models we considered with the methods we used. It would be interesting to investigate other quantum algorithms for ground-state preparation on these models such as AVQITE~\cite{Gomes:2021ckn} 
and compare with the ADAPT-VQE results presented here. We also found that the sparse SYK Hamiltonian with $k_s = 9$ we considered is not sparse in the sense of the Hamiltonians considered in 
Ref.~\cite{Chen2024}. We provide more details about this in Appendix~\ref{app:DSH}. So, it remains an open question as to whether ground-state preparation can be efficient for these models on a quantum computer for less-sparse models ($k_s \gg 1$). We also tested the transferability of the ansatz, i.e., whether results obtained from training on a particular instance of the disordered Hamiltonian can be used to train other instances with reduced computational resources. Details are given in Appendix~\ref{app:ToA}, where the results suggest that this does not work in practice. In this regard, another option is to pursue the ideas of eigenvector continuation~\cite{Frame:2017fah, Agrawal:2024xjs} and construct ground states for a few instances and then use them to efficiently compute the disorder averages by building appropriate subspaces. We leave this for future work.\\ 

\section*{Acknowledgments}
S.G acknowledges the support of the Deutsche Forschungsgemeinschaft (DFG, German Research Foundation) under Germany's Excellence Strategy - EXC-2123 Quantum-Frontiers - 390837967. We acknowledge support by the U.S. Department of Energy, Office of Science, Office of Advanced Scientific Computing Research under Award Number DE-SC0025430 that supported the ADAPT-VQE simulations, and Award Number DE-SC0025384 which supported the conception, initial planning and writing of the project.

\section*{Data Availability}
The code and data can be obtained from the authors on reasonable request. 


\bibliographystyle{utphys}

\bibliography{v1.bib}


\appendix

\section{\label{sec:AVQE}ADAPT-VQE}
In this section, we review the ADAPT-VQE algorithm that we employ in this work~\cite{Grimsley:2018wnd, Grimsley:2022azc}. Unlike VQEs that use a fixed ansatz structure, ADAPT-VQE builds it dynamically from a pool of operators chosen depending on the problem at hand. The operator (or operators~\cite{AnastasiouPRR2024}) to add to the ansatz at each adaptive iteration is chosen based on their local gradients, which can be measured efficiently on quantum computers. In other words, only operators that change the energy beyond some threshold are added to the ansatz. After each adaptive iteration, all the parameters in the current ansatz are optimized using a VQE subroutine, which involves a feedback loop between energy and optimization gradient measurements (if necessary) on the quantum hardware, and parameter updates proposed by a classical optimizer. After $n$ iterations of ADAPT-VQE, the parametrized quantum state is given by:
\begin{equation}
    \ket{\psi^{(n)}(\bm{\theta})} = e^{-i\,\theta_n \hat{A}_n} \cdots e^{-i\,\theta_1 \hat{A}_1} \ket{\psi_{0}},
    \label{eq:state_adapt}
\end{equation}
where $\ket{\psi_{0}}$ is the reference state we start with. This is typically chosen as some physically motivated state that is easy to prepare. The operators $A_i$ are Hermitian generators, drawn from the operator pool. The local gradients of these operators that determine their selection can be measured efficiently:
\begin{equation}
    \frac{\partial \langle H \rangle^{(n)}}{\partial \theta_{n+1}}\Bigg\vert_{\theta_{n+1} \to 0}= i\,\langle{\psi^{(n)}} \vert [H, \hat{A}_{n+1}] \vert \psi^{(n)}\rangle,
\end{equation}
where $\ket{\psi^{(n)}}$ is the ADAPT-VQE state defined in Eq.~\eqref{eq:state_adapt}. The ADAPT-VQE algorithm terminates when certain convergence criteria are met. This typically is a threshold on the norm of the gradient vector of the pool operators. This problem-tailored approach to constructing the ansatz helps avoid optimization problems such as local traps and barren plateaus that are encountered in some VQEs with fixed ans\"{a}tze~\cite{Grimsley:2022azc}. ADAPT-VQE has been numerically shown to outperform VQEs with fixed ans\"{a}tze~\cite{Ramoa:2024zed} in terms of quantum and classical resources.

\section{\label{app:ToA}Transferability of ADAPT ansatz}

All the models we have considered in this work belong to the class of random Hamiltonians. Usually, in these cases one has to average over multiple instances of the model. In this section, we explore the possibility of transferring the ADAPT-generated variational ansatz trained on a specific instance to other instances of the model without going through the optimization process again. We performed a transferability analysis across independent disorder realizations of the SYK Hamiltonian at fixed system size $N$. For each $N$, we first obtain a converged ADAPT-VQE ansatz for a given disorder instance, yielding a sequence of operator indices $\{\hat{A}_k\}$ and optimized parameters $\{\theta_k\}$. To test transferability, we applied the same ansatz to a new disorder realization at the same $N_f$, but re-optimized all variational parameters through a classical optimizer before adding any more operators needed for the new instance.  After parameter re-optimization, we resumed the ADAPT procedure starting from the transferred ansatz and monitored whether additional operators were selected by the gradient criterion. 

For small system sizes ($N=10$ and $N=12$), the transferred ans\"atze remained highly effective after re-optimization: the re-optimized transferred circuits for 5 different instances for each $N$ achieved relative energy errors in the range of $0.003\%$--$0.005\%$, with ground-state fidelities consistently reaching $\sim 0.9999$ and only a very few additional operators were required upon resuming ADAPT. This indicates that, in the small-$N$ regime, the operator sequence identified by ADAPT captures structural features that are largely independent of the specific disorder realization.

In contrast, for larger system sizes ($N=14$ and $N=16$), the performance of the transferability scheme mentioned above reduces. Although parameter re-optimization lowered the energy compared to the raw transferred state, the fidelity remained poor in many cases. Upon resuming ADAPT, a substantial number of new operators were selected, often comparable to the size of a fresh ADAPT run. This behavior indicates that, at larger $N$, the optimal operator sequence becomes strongly disorder-dependent.

These results suggest that while ADAPT identifies partially universal operator structures in small systems, the variational manifold becomes increasingly instance-specific as system size grows, reflecting the enhanced complexity of the SYK ground state in the large $N$ regime. This behavior is physically expected: independent disorder realizations of the SYK model can differ substantially in their microscopic coupling structure, and the associated ground states need not admit a simple shared operator decomposition. As $N$ increases and the many-body Hilbert space grows exponentially, the variational manifold identified by ADAPT naturally becomes more sensitive to the specific realization of disorder. It might be useful to pursue ideas of eigenvector continuation for disordered systems to achieve an advantage when averaging over instances. 

\section{\label{app:presence_BP}Presence of barren plateau}

We now address the issue of occurrence of barren plateau in the models we have considered. Barren plateau is a variational problem~\cite{Larocca:2024plh} where the gradient of the cost function concentrates around zero leading to exponential increase in the optimization process. Recently, the issue of barren plateau was argued to be related to the exponential size of the dynamical Lie algebra (DLA) of a given Hamiltonian~\cite{Ragone:2023qbn}. For variational preparation of the ground states, the dimension of the DLA is directly related to the barren plateau problem where the variance of the loss function concentrates around zero and is inversely related to the size of the dynamical Lie algebra (DLA) of the problem Hamiltonian. The dimension of the DLA for both dense and the sparse SYK model (with $k=9$) scales exponentially with $N$. In particular, we find that $\text{dim}\,(\mathfrak{g}) = 2^{2n-1}-2 = 2(2^{2n-2} - 1) = \oplus_{i=1}^{2} \mathfrak{su}(2^{n-1})$ where $n= N/2$ is the number of qubits for the dense SYK model while
for the SK model, we find that 
$\text{dim}\,(\mathfrak{g}) = 2(4^{L-1}-1) = \mathfrak{su}(2^{L-1}) \, \oplus \mathfrak{su}(2^{L-1})$ where $L$ is the number of qubits. 

Our results clearly show that while the optimization is well-behaved and scales polynomially for the SK model, it is exponentially hard for the dense and sparse SYK models. This captures that the exponential dimension of the DLA does not necessarily fully correlated with barren plateau for all cases. This leads to the deeper question of what actually determines the efficacy of the variational algorithms given a Hamiltonian. In this regard, some notable differences between the SK model and SYK models is that the former has a much larger spectral gap and is stoquastic~\cite{Bravyi:2007gix} which implies that it does not suffer from the sign problem while SYK has a smaller gap (see Fig.~\ref{fig:syk_exact_gs_gap}) and is not stoquastic leading to a severe sign problem. 
In addition, the SK model is also $O(1)$ $d$-sparse unlike the SYK Hamiltonian as discussed in Appendix~\ref{app:DSH}. It would be interesting to understand the connection between sparsification of the SYK model and the severeness of the sign problem. We leave this for future work. It would also clarify the quantum advantage in simulating many-body systems with strong sign problem.

\section{\label{app:DSH}$d$-sparse Hamiltonians}

In this paper, we use the term `sparse SYK' to denote the simplification of the Hamiltonian where we only keep $k_s N$ terms in the Hamiltonian with $k_s = 9$. This notion of sparsity is different from that considered in Refs.~\cite{Chen2024, Herasymenko:2022wup} where it is in terms of $d$-sparseness of the Hamiltonian. While $d$-sparse Hamiltonians always implies sparse SYK model, the opposite is not always true. In Table.~\ref{tab:syk_dense_sparse_d}, we show the $d$-sparsity of the sparse SYK Hamiltonian which we have considered in this paper. 

\renewcommand{\arraystretch}{1.4} 
\begin{table}[h!]
\centering
\begin{tabular}{c|c|c}
\toprule
$n = N/2$ & Dense SYK & Sparse-SYK, $k_s=9$ \\ \hline
8 & 99 & 76 \\ \hline
9 & 163 & 103 \\ \hline
10 & 256 & 127 \\ \hline
12 & 562 & 178 \\ \hline
13 & 794 & 200 \\ \hline
\end{tabular}
\caption{Comparison of the $d$-sparsity of the dense and sparse ($k=9$) SYK models for different number of qubits, $n = N/2$. The results are averaged over 5 instances and rounded to the nearest integer for the sparse model. For the SK model, $d$-sparsity is $n+1$.}
\label{tab:syk_dense_sparse_d}
\end{table}

\section{Reference state choice}
\label{app:robustness}

To address the question of whether the inefficient scaling of ADAPT-VQE resources observed for the SYK model is a consequence of poor choice of reference state or operator pool, we performed additional numerical tests for $N = 10, 12, 14$ Majorana fermions, averaging over three disorder instances each for both the dense and sparse ($k_s = 9$) models.

The first reference state we tested follows the 
Clifford Ansatz For Quantum
Accuracy (CAFQA) ~\cite{Ravi2023CAFQA} procedure adapted for SYK. Starting from the N\'{e}el state $|01\rangle^{\otimes n/2}$ in the correct $\mathbb{Z}_2$ parity sector, we apply $\tilde{L}$ layers of particle-number-preserving iSWAP gates $\exp\left(-i\frac{\pi}{4}(X_iX_j + Y_iY_j)\right)$ to generate a Clifford circuit ansatz. For each candidate circuit, the energy $\langle\psi_{\rm Clifford}|H|\psi_{\rm Clifford}\rangle$ is evaluated classically using the stabilizer tableau formalism, which is efficient since Clifford circuits are classically simulable via the Gottesman-Knill theorem~\cite{gottesman1997}. We perform an exhaustive search over all possible pair sequences for $\tilde{L} = 2$ layers, corresponding to search spaces of size 100, 225, and 441 for $N = 10, 12, 14$ respectively, and select the circuit minimizing the energy for each disorder instance. This constitutes a Hamiltonian-specific Clifford warm start requiring no variational optimization. For the dense SYK model, the optimal Clifford reference state did not reduce the number of variational parameters or circuit depth compared to the N\'{e}el baseline. For the sparse SYK model, some improvement in convergence reliability was observed, though at comparable circuit cost. We conclude that even a fully optimized Clifford warm start using the CAFQA protocol does not reduce the inefficient resource scaling ground state preparation of the SYK model using ADAPT-VQE.

Second, we construct a reference state following the stabilizer configuration interaction (SCI)~\cite{SCI2025} approach, adapted for SYK. In SCI, one starts from a reference stabilizer state and iteratively applies excitation operators, $\mathcal{E}$, of the form $X_{\{o\}}X_{\{u\}}$, acting on occupied and unoccupied orbitals respectively, as $|\psi_{i+1}\rangle = (\mathbb{I} + (-1)^l \mathcal{E}_{i+1})|\psi_i\rangle/\sqrt{2}$ with $l = 0, 1$, preserving the stabilizer nature of the state at each step. The resulting stabilizer state with the lowest energy is selected as the reference state for ADAPT-VQE. For SYK, we start with the N\'{e}el state $|01\rangle^{\otimes n/2}$ in the correct $\mathbb{Z}_2$ parity sector, and particle-number-preserving $X_oX_u$ operators acting on qubit pairs as the excitation generators. We enumerate all valid excitation sets and evaluate the energy of each resulting stabilizer state classically, selecting the one with lowest energy for each disorder instance. However, since the SYK ground state exhibits volume-law entanglement, no stabilizer state can serve as a good approximation. We observe that the SCI reference state achieves only a modest overlap with the true ground state and does not reduce the circuit resources required for ADAPT-VQE to converge.

\end{document}